\documentclass[manuscript]{aastex}
\usepackage{natbib}
\shorttitle{Frerequencies of Magnetic Processes on Supergranule Boundary}
\shortauthors{Y. Iida et al.}

\begin{document}

\title{Detection of flux emergence, splitting, merging, and cancellation of network field. I Splitting and Merging}

\author{Y. Iida}
\affil{Department of Earth and Planetary Science, University of Tokyo, Hongo, Bunkyo-ku, Tokyo, 113-0033, Japan}

\author{H. J. Hagenaar}
\affil{Lockheed Martin Advanced Technology Center, Org. ADBS, Building 252, 3251 Hanover Street, Palo Alto, CA 94304}

\and

\author{T. Yokoyama}
\affil{Department of Earth and Planetary Science, University of Tokyo, Hongo, Bunkyo-ku, Tokyo, 113-0033, Japan}

\begin{abstract}
Frequencies of magnetic patch processes on supergranule boundary, namely flux emergence, splitting, merging, and 
cancellation, are investigated through an automatic detection.
We use a set of line of sight magnetograms taken by the Solar Optical Telescope (SOT) on board {\it Hinode} satellite.
We found 1636 positive patches and 1637 negative patches in the data set, whose time duration is 3.5 hours and field of view is 
$112" \times 112"$.
Total numbers of magnetic processes are followed: 493 positive and 482 negative splittings, 536 positive and 535 negative 
mergings, 86 cancellations, and 3 emergences.
Total numbers of emergence and cancellation are significantly smaller than those of splitting and merging.
Further, frequency dependences of merging and splitting processes on flux content are investigated.
Merging has a weak dependence on flux content only with a power-law index of $0.28$.
Timescale for splitting is found to be independent of parent flux content before splitting, which corresponds to $\thicksim$33 minutes.
It is also found that patches split into any flux contents with a same probability.
This splitting has a power-law distribution of flux content with an index of $-2$ as a time independent solution.
These results support that the frequency distribution of flux content in the analyzed flux range is rapidly maintained by merging and splitting, 
namely surface processes. 
We suggest a model for frequency distributions of cancellation and emergence based on this idea.

\end{abstract}


\section{Introduction}

How magnetic structure on the solar surface is constructed and maintained is one of the fundamental issues in solar magnetic field observation. 
It is important for the statistical understanding of solar activities because they are triggered by magnetic activities on the solar surface. 
It may also give a quantitative restriction to solar dynamo problem.
One approach is to investigate a frequency distribution of magnetic flux content on the solar surface.
Some authors found an exponential distribution of flux content \citep{sch1997b,hag1999}.
On the other hand, other authors found a power-law distribution \citep{wan1995, par2009, zha2010}.
\cite{par2009} reported a power-law distribution with an index of $-1.85$ between $2 \times 10^{17}$ Mx and $10^{23}$ Mx,
 which means that magnetic patches from large active regions to small patches in quiet network are described
 by a single flux distribution.
They suggest the idea that either all surface magnetic features are generated by the same mechanism or that they are dominated by the surface processes.

The next arising question is how the frequency distribution is achieved and sustained.
Magnetic processes, namely flux emergence, splitting, merging, and cancellation of magnetic patches on the photosphere are
 thought to change and maintain the frequency distribution.
Relationship between these magnetic processes and the flux distribution is described by magneto-chemistry equation \citep{sch1997b}.
Based on this equation and some assumptions, they found a time independent solution for an exponential frequency distribution of flux content.
Furthermore \cite{par2002} found a particular solution for an arbitrary frequency distribution of flux content.
Both solutions assume re-appearances of canceled fluxes and a detailed balance between any two fluxes,
 which should be verified observationally.

From the view of flux balance on the photosphere, flux emergence and cancellation are especially investigated because
 they have a direct relation with a flux exchange through the photosphere. 
Flux emergence, which is observed as a divergence of opposite polarities in magnetogram, is thought to be a flux ascension from 
below the photosphere. It produces flux increases of both polarities in line-of-sight magnetograms.
The frequency distribution of emerging flux is investigated by several authors.
\cite{hag2001} found an exponential distribution by using full disk magnetograms obtained by SOHO/MDI.
\cite{tho2011} found a power-law distribution from active region to inter-network field by using {\it Hinode}/SOT magnetogram.
Flux cancellation is defined as a convergence and a disappearance of magnetic fluxes of positive and negative polarities in 
line-of-sight magnetograms \citep{mar1985,liv1985}.
It decreases both positive and negative fluxes on the solar surface.
Two physical models, U-loop emergence and $\Omega$-loop submergence, are proposed \citep{zwa1987} and many authors tried to 
distinguish them \citep{har1999, yur2001, cha2002, kub2010, iid2010, cha2010}.
Flux replacement timescales by emergence and cancellation are physical quantities representing importance of these processes 
on the solar surface.
But it varies from several days to several hours \citep{liv1985, sch1998, hag2001} and it is not clear why this timescale decreases
as the spatial resolution becomes higher.

Merging and splitting should also take important roles in flux maintenance because magnetic patches change their flux content
 through these processes.
The action of convective motion should differ among patches with different flux content. 
However, there are fewer reports for merging and splitting compared to those of emergence and cancellation.
The role of these processes in flux balance still remains unclear.

We investigate the frequencies of magnetic patch processes, namely flux emergence, splitting, merging, and cancellation
 based on observations in this series of papers.
 This first paper is mainly devoted to investigations of total flux amount of magnetic processes and frequency distributions 
on flux content of merging and splitting.
The purpose of this paper is to clarify what the processes dominate the frequency distribution of flux content.
Our final goal of this series of papers is to understand how the flux distribution is maintained on the solar surface.
We concentrate magnetic processes on the network because most flux is contained there \citep{mar1990,sch1998}.
The detail of data set is explained in section 2.
We explain our definition of a magnetic patch and magnetic processes in section 3.
The results are shown in section 4 and related discussions in section 5.

\section{Observation}

We use line-of-sight magnetograms near the disk center obtained by Narrowband Filter Imager (NFI) of Solar Optical Telescope (SOT) 
onboard {\it Hinode} satellite \citep{kos2007, tsu2008, ich2008, shi2008, sue2008}.
We need to use {\it Hinode}/NFI magnetograms in this study because the detection limit of MDI high-resolution data is $\thicksim 10^{18}$ Mx 
(see a Figure 5 (a) by \cite{par2009}) and we can not set a flux range wide enough for an analysis of the network field.  
Time period of the data set is from 0:33UT to 4:08UT on 2009 November 11.
NFI observed magnetograms near disk center by using Na I D$_1$ resonance line at $5896 \ \AA$ during this period.
Full field of view is $112'' \times 112''$.
Figure 1 shows an example of magnetograms used in this study after the pre-process explained in the next section.
Some network cells, which has a typical size of $20-40''$, are included in the field of view.
Total number of magnetograms is 199 during the whole observational period.
Time interval between each magnetogram is 1 minute.
The interval is sometimes 2 minutes due to the lack of data.
The region is a quiet Sun because there are no active regions on the solar disk during the observational period.

Some corrections of SOT magnetogram are done before detecting magnetic processes.
We use $\tt fg\_prep.pro$ procedure in SolarSoftWare (SSW) package for a correction of dark current and flat field of CCD camera.
The data is rotated to the position at November 11 2:03UT when the region is near the disk center 
by using $\tt drot\_map.pro$ procedure in SSW package.
We remove a columnwise median offset of CCD camera \citep{lam2010} by subtracting a median value of 
each column from all pixels in the same column.
Small spatial fluctuation of field of view is removed by making a correlation between consecutive images.
Then the magnetograms are averaged over 3 continuous images and 3 pixels for smoothing.

The observed circular polarization signals (CP) are converted to the line-of-site magnetic field strength by the following method.
We multiply CP by a conversion coefficient to obtain actual magnetic field strength in the next step.
This conversion coefficient is determined by comparing the SoHO/MDI HR magnetogram and CP image on 2009 November 11 1:39UT.
We spatially smear the SOT data to two times of MDI pixel size (R. A. Shine in private communication) and make a linear fitting
 between CP in SOT and magnetic field in MDI.
Because there are too weak or strong pixels in SOT data which are out of linear range, we make the fitting in the range 
from $30$ G to $100$ G as a magnetic field strength.
Figure 2 shows a scatter plot of SOT data number and magnetic field strength obtained by MDI.
The solid line shows a result of linear fitting.
We obtain $9067.98$ G/DN as a conversion coefficient.

\section{Identification of Magnetic Processes}

We define magnetic processes and explain our method for the identification of them in this section.
Figure 3 represents schematic pictures of four magnetic processes detected in this study and 
Figure 4 summarizes our definition of magnetic processes.
Note that our definitions of these processes are valid even when more than two patches are involved in the process 
but invalid when processes which increase and decrease flux content occur at the same time.

\subsection{Detection and Tracking of Magnetic Patches}
We use a clumping method for a detection of magnetic patches: 
each patch is picked up as a clump of marked pixels having magnetic strength beyond a given threshold \citep{par2002}.
The adopted threshold is one sigma obtained by fitting the histogram of the signed magnetic strength in each magnetogram,
 i.e. $\sigma \thicksim 5$ G pixel$\ \thicksim 6.8 \times 10^{14}$ Mx.
This value is close to that of \cite{par2009}.
A magnetic patch in a clumping method corresponds to a massif of magnetic patches in a downhill method and a curvature method which 
are used in some previous studies \citep{hag1999b, wel2003, def2007}.
We pick up magnetic patches with sizes beyond 81 pixels for focusing our analysis on the network magnetic field. 
The validity of this choice is demonstrated in Figure 5, in which the network structure is seen when adopting the 81-pixels threshold.

We track the motion of magnetic patches in consecutive images after the detection of them.
Patches are marked as identical when they have a spatial overlapping in continuous images \citep{hag1999}.
It should be noted that a travel distance of magnetic patches in the data interval ($\thicksim$1min) is nearly up to 
one pixel size, namely $2 \ \rm{km}\ \rm{s}^{-1}\ \times$ 1 $\rm{min} = 120\ \rm{km} \thicksim 1$ pixel. 
In high-resolution magnetogram, more than one patch in a previous image often have spatial overlappings with one patch
 in a consecutive image and vice versa.
To clear up this problem, we set two conditions when tracking patches.
First we investigate spatial overlappings of patches from those with larger flux contents.
This is based on a concept that a smaller patch has a greater tendency to fall below a detection limit of the analysis 
by splitting and cancellation.
Second we select a patch with the most proximate flux content in case of overlappings of more than one patch.
Tracking paths of detected patches become unique with these conditions.

\subsection{Merging and Splitting}

Merging is a process where more than one patch of same polarity converge and coalesce to one patch (Figure 3a).
We define a merging as an event with two conditions.
A merging event is defined as a feature which satisfies the conditions 
1) that there are two or more parent patches in a previous magnetogram overlapping one daughter patch in the consecutive magnetogram,
2) and that more than one of the concerned patches in the previous magnetogram disappear in the time interval.
Panels in the first row of Figure 6 show an example of detected mergings.
Two patches converge in the first and second images.
They coalesce to one patch in the third image.

Splitting is a process where a single patch is divided into more than one patch (Figure 3b).
A splitting event is defined as a feature which satisfies the two conditions
1) that there are one or more daughter patches in a magnetogram overlapping one parent patch in the previous magnetogram,
2) and that more than one of the concerned patches in the latter magnetogram appear in the time interval.
Panels in the second row of Figure 6 show an example of detected splittings.
One patch stays in the first and the second panel.
It splits into two patches between the second and third panel.

\subsection{Emergence and Cancellation}

The detailed explanation for the detection technique of emergence and cancellation will be given in the next paper 
in which much more events are detected by using another data set, making it possible for us to conduct a statistical study of them.
Emergence and cancellation are defined as a pair of flux change events of each patch in opposite polarities within a certain distance.
We pick up not only complete cancellations but also partial cancellations in this definition.

Panels in the third row of Figure 6 show an example of emergences.
In the second panel, positive and negative patches appear but the positive patch is already coalesced with a pre-existing patch.
These emerging patches continue to separate against each other and become distant in the last panel.
Panels in the last row of Figure 6 show an example of partial cancellations.
Opposite polarities are converging in the first and second panel.
They start canceling between the second and third panel.
The negative patch totally disappears in the last panel but positive patch remains after the cancellation in this example.

\section{Results}

\subsection{General Description}
Along with the total numbers of patches and their flux density averaged over field of view and the observational period (in Mx cm$^{-2}$) and 
the frequencies of events in magnetic flux density (in Mx cm$^{-2}$ s$^{-1}$) are summarized in Table 1.
Total numbers of detected negative and positive patches are $1636$ and $1637$, respectively, enough for a statistical study.
As for magnetic processes, merging and splitting are much more frequent than emergence and cancellation.
Total numbers of merging for positive and negative patches are $493$ and $482$.
Total numbers of splitting for positive and negative patches are $536$ and $535$.
These are enough for a statistical study.
The number of emergence is $3$ and that of cancellation is $86$, which are not enough for a statistical study.

To determine what magnetic processes dominate a flux balance in this region, we investigate total flux amounts of patches and those related to 
the magnetic processes per unit area per unit time.
The results are shown in Table 1.
The averaged flux density are $2.53$ Mx cm$^{-2}$ and $3.60$ Mx cm$^{-2}$ for positive and negative patches respectively.
They correspond to the total flux amount of $1.63 \ \times \ 10^{20}$ Mx and $2.32 \ \times \ 10^{20}$ Mx for our field of view.
The rate of averaged flux density involved in merging processes is $1.65 \ (3.53) \times \  10^{-3}$ Mx cm$^{-2}$ s$^{-1}$ for positive (negative) patches.
We define flux amount of merging process as that of parent patches.
Flux replacement timescale by positive (negative) merging, $\tau_{\rm mrg}=\Phi_{\rm tot}/ ( \left. \partial \Phi_{\rm tot}/\partial t \right|_{\rm mrg})$, is evaluated as 
$1.53 \ (1.02) \times 10^3$ sec from these values, which is much shorter than the estimated replacement timescale by cancellation and emergence reported in 
the previous studies \citep{liv1985, sch1998, hag2001, tho2011}.  
The rate of averaged flux density involved in splitting processes is $1.48 \ (3.03) \ \times \ 10^{-3}$ Mx cm$^{-2}$ s$^{-1}$ for positive (negative) patches.
We define flux amount of splitting process as sum of those of parent patches.
In the same manner as merging, flux replacement timescale by positive (negative) splitting is evaluated as $1.71 \ (1.19) \times 10^3$ sec.

\subsection{Frequency distribution of flux content}

Figure 7 shows a frequency distribution of flux content.
Red and blue dashed histogram denote those of positive and negative polarities, respectively. 
Solid histogram denotes sum of them. Black dashed line indicates the fitting result of it. 
The distribution is dropping down below $\phi_{\rm{th}} = 10^{17.5} \ \rm{Mx}$, which is suggested as a detection limit in this study, 
below which only a limited number of the patches are detected. 
Dashed line is a fitted power-law result between $10^{17.5} \ \rm{Mx}$ and $10^{19} \ \rm{Mx}$.
We make a least-square fitting of power-law function with a form, $n(\phi)=n_0 ( \phi / \phi_0 )^{-\gamma}$. 
We obtain $n_0 = 2.42 \times 10^{-36}$ Mx$^{-1}$ cm$^{-2}$, $\phi_0 =1.0 \times 10^{18}$ Mx, and $\gamma = 1.78 \pm  0.05$.
The error means one sigma error.
The fitted power-law index is sensitive to fitting range as reported by \cite{par2009}.
The index varies from $-1.78$ to $-1.91$ with a minimum fitting range of $10^{17.5}$ to $10^{18}$ Mx.
Total error of our fitting of the power-law index becomes $0.18$ with considering this error.
This result is consistent with \cite{par2009}, which reports $-1.85 \pm 0.14$ as an power-law index 
of a frequency distribution of flux content. 

\subsection{Probability distributions on flux content of merging and splitting}

We investigate frequency distributions on flux content of merging and splitting for one patch, 
which are defined as frequency distributions of processes divided by a frequency distribution of flux content.
The value of these distributions represent probability of the processes.
Further, the detection limit $\phi_{\rm th}$ is taken into the account in this study. 
We call them apparent probability disributions of processes.

Figure 8 shows the apparent probability distribution of merging.
We make a least-square fitting in a range of $10^{17.5}$ - $10^{19}$ Mx, where number of merging event is enough for fitting.
The fitting form is 
\begin{equation}
\frac{\partial P_{\rm mrg}^{\rm APP}}{\partial t}=p_{0,{\rm mrg}} \left( \frac{\phi}{\phi_0} \right)^{\beta_{\rm mrg}}
\end{equation}
where $p_{0,{\rm mrg}}$ is a reference probability, $\phi_0$ is a reference flux content, and $\beta_{\rm mrg}$ is a power-law index of
probability distribution of merging.
We obtained $p_{0,{\rm mrg}}=(2.52\pm0.08) \times 10^{-4}$ s$^{-1}$ and $\beta_{\rm mrg}=0.28\pm0.05$ for positive patches, 
$p_{0,{\rm mrg}}=(2.52.\pm0.08) \times 10^{-4}$ s$^{-1}$ and $\beta_{\rm mrg}=0.26\pm0.05$ for negative patches, 
and $p_{0,{\rm mrg}}=(5.12\pm0.11) \times 10^{-4}$ s$^{-1}$ and $\beta_{\rm mrg}=0.28\pm0.04$ for both patches with $\phi_0=10^{18}$ Mx.
The errors mean one sigma error of least-square fitting.

Figure 9 shows the apparent probability distribution of splitting.
The strong increase in the range larger than $10^{19}$ Mx is caused by lack of the patch number in the analysis.
On the other hand, there is a drop in a flux range near $\phi_{\rm{th}}$ where the number of patches is enough for a statistical study. 
We interpret this dropping as an effect of splitting into the area below $\phi_{\rm{th}}$. 
This effect is evaluated in the discussion in Section 5.1. 
We see that probability of splitting is almost constant as $1.0 \times 10^{-3} \  \rm{sec}^{-1}$, which means a timescale of $33$ minutes, 
in the range enough above $\phi_{\rm{th}}$, $3.0 \times 10^{18} - 1.0 \times 10^{19}$ Mx.
It means that frequency of splitting is independent of parents' flux content.

\section{Discussion}

\subsection{Probability density distributions of merging and splitting}

We evaluate probability density distributions of merging and splitting terms in magneto-chemistry equation from the observational results.
Because probability distributions, which we obtained in this study, are obtained by integrating them on flux content once, 
we have to put at least one assumption to evaluate them.

The merging function $l(x,y)$ is given as follows. 
We obtain the form as a probability distribution from the definition of $l(x,y)$ as
\begin{equation}
\frac{\partial P_{\rm mrg}^{\rm APP}}{\partial t}=\int_{\phi_{\rm th}}^{\infty}n(x)l(\phi,x)\, d\phi .
\end{equation}
By comparing with the observational result, we obtain  
\begin{equation}
\displaystyle \int_{\phi_{\rm th}}^{\infty}n(x)l(\phi,x)dx=p_{0,{\rm mrg}}\left( \frac{\phi}{\phi_0} \right)^{\beta_{\rm mrg}} 
\ \ \ (\phi \ge \phi_{\rm th}). \label{eq:rel_mrg}
\end{equation}
In the left-hand-side of this equation, the variable $\phi$ appears only in the $l(\phi,x)$.
We assume a simple form satisfying this relation namely,
\begin{equation}
l(x,y) \propto x^{\beta_{\rm mrg}}.
\end{equation}
From the symmetry of $l(x,y)=l(y,x)$, this relation deduces 
\begin{equation}
l(x,y) = l_0 \left( \frac{x}{\phi_0} \right)^{\beta_{\rm mrg}} \left( \frac{y}{\phi_0} \right)^{\beta_{\rm mrg}}.
\end{equation}
Substituting it and observational result of $n(\phi)$ into Eq.(3), we obtain
\begin{equation}
\displaystyle l_0=\frac{\displaystyle p_{0,{\rm mrg}}}{\displaystyle n_0 \phi_0^{\gamma-\beta_{\rm mrg}} \int_{\phi_{\rm th}}^{\infty}x^{-\gamma+\beta_{\rm mrg}}dx}.
\end{equation}
The upper value of the integration is limited on the actual Sun and we put it as $\phi_{\rm max}$.
We substitute the value obtained in our study, namely $n_0=1.21\times10^{-36}$ Mx$^{-1}$ cm$^{-2}$, $\gamma=1.78$, $p_{0,{\rm mrg}}=2.56\times10^{-4}$ s$^{-1}$, 
$\beta_{\rm mrg}=0.28$, $\phi_{\rm max}=10^{19}$ Mx, $\phi_{\rm th}=10^{17.5}$ Mx, and $\phi_0=10^{18}$ Mx, and
 obtain $l_0=7.24 \times 10^{13}$ cm$^2$ s$^{-1}$.

The splitting function $k(x,y)$ is given as follows. 
From our observations, the probability distribution of splitting events $\partial P^{\rm APP}_{\rm splt}(\phi) / \partial t$ is suggested to be independent
of the parent patch flux:
\begin{equation}
\frac{\partial P^{\rm APP}_{\rm splt}}{\partial t}(\phi)=k_0=\rm{constant} \ (\rm{for \ all} \ \it{\phi}).
\end{equation}
This claim is observationally supported at least in the range $\phi > \phi_{\rm{th}}$ (Figure 9).
The drop off below $\phi_{\rm{th}}$ is discussed immediately below.
If the splitting ratio between daughter patches is randomly determined, i.e.
\begin{equation}
\frac{\partial}{\partial x} \left[ k(x,\phi-x) \right] = 0 \ (\rm{for} \ 0 < \it{x} < \phi),
\end{equation}
then we obtain,
\begin{equation}
k(x,y)=\frac{k_0}{x+y}.
\end{equation}
When the flux content of the daughter patch is below $\phi_{\rm{th}}$, such events are {\it not} recognized as a splitting event in our procedure. 
The probability distribution will be given as
\begin{equation}
\frac{\partial P^{\rm APP}_{\rm splt}}{\partial t}(\phi) = \int_{\phi_{\rm{th}}}^{\phi-\phi_{\rm{th}}}k(x,\phi-x)\, dx = k_0 (1-\frac{2 \phi_{\rm{th}}}{\phi}).
\end{equation}
Black and blue dashed curves in Figure 9 indicate analytical curves with $k_0=1.0\times10^{-3}$ s$^{-1}$ and $k_0=5.0\times10^{-4}$ s$^{-1}$ 
respectively.
This curve fits the drop of the observational line well, which supports the above assumptions.
We obtain $k(x,y)=k_0/(x+y)$ as splitting function, where $k_0=5.0\times10^{-4}$ s$^{-1}$.

\subsection{The time-independent solution of splitting process}
Since our observations show that merging and splitting are much more frequent than emergence and cancellation, it suggests that the former 
two have much influence on the maintenance on the power-law distribution. 
We show the time-independent solution by splitting although we have not found the time-independent solution by merging and splitting.
The magneto-chemistry equation is consist of source (emergence) term, merging terms, splitting terms, and cancellation terms 
(see the right-hand-side of Eq.(3) in \cite{sch1997b}). 
The frequency of emergence, merge, splitting, and cancellation are represented by $S(\phi)$, $l(x,y)$, $k(x,y)$, and $m(x,y)$ respectively.
We use the magneto-chemistry equation only including the splitting terms by setting $S(\phi)=0$, $l(x,y)=0$, $m(x,y)=0$, namely 
\begin{equation}
\frac{\partial n(\phi)}{\partial t} = 2\int_{\phi}^{\infty}n(x)k(\phi,x-\phi)\, dx - \int_{0}^{\phi}n(\phi)k(x,\phi-x)\, dx
\end{equation}
where $n(\phi)$ is a frequency distribution of flux content.\footnote{Terms in the right-handside of this equation is 
different from those of Eq.(3) in \cite{sch1997b} at the point of the splitting from $\phi$, namely $n(x)k(\phi,x-\phi)$ in this paper 
and $n(x)k(\phi,x)$ in \cite{sch1997b}. The term in \cite{sch1997b} should be a typo because we have to multiply the pre-splitting 
number density here.}

After substituting $k(x,y)=k_0/(x+y)$ into Eq.(11) and differentiating with $\phi$, we obtain
\begin{equation}
\frac{\partial^2n(\phi)}{\partial \phi \partial t} = - \frac{k_0}{\phi^2}\frac{\partial}{\partial \phi}\left[\phi^2n(\phi)\right].
\end{equation}
This equation has a time-independent solution $n(\phi) \propto \phi^{-2}$.
This power-law index of flux content is in good agreement with the observational result.
This scale free distribution comes from constancies of splitting, namely that splitting has a constant timescale independent of flux content 
and constant probability of splitting to flux content.
These constancies may come from convection dominating patch stability or flux tube instability with a constant timescale.
We needs further theoretical and observational studies to determine which of the hypotheses is the actual 
scenario on the solar surface.

\subsection{Relationship among frequency distribution of flux content, cancellation, and emergence}

We suggest a model of relationship among frequency distribution of flux content, cancellation, and emergence.
The important hypothesis in this model is that the frequency distribution of flux content is rapidly maintained regardless of cancellation and emergence,
which is supported by this study.
Figure 10 shows a schematic view of this model.
We put a power-law distribution of flux content in unit of patches Mx$^{-1}$ cm$^{-2}$ as 
\begin{equation}
n(\phi)=n_0 \left( \frac{\phi}{\phi_0} \right)^{-\gamma}
\end{equation}
where $\phi_0$ is a reference value of flux content and $n_0$ is a reference frequency of flux content.
The power-law index, $\gamma$, is derived as $1.5<\gamma<2$ by our observation and the previous studies \citep{par2009,zha2010}.
The maximum flux content in the system ($\phi_{\rm{max}}$) is assumed to be much larger than the minimum ($\phi_{\rm{min}}$) in the following discussion.
We calculate $N(\phi)$, a total patch number with flux content larger than $\phi$, by integrating a flux distribution
 from $\phi$ to $\phi_{\rm max}$ as
\begin{equation}
N(\phi)=\int_{\phi}^{\phi_{\rm max}}n(\phi^{\prime}) \, d\phi^{\prime} \approx \frac{n_0 \phi_0}{\gamma-1} 
\left( \frac{\phi}{\phi_0} \right)^{-\gamma+1}.
\end{equation}
\cite{sch1997b} evaluated a collision rate of opposite patches from a total patch number density with assumptions of a 
constant velocity and a randomness of patch motion along network.
They obtained the collision frequency, $\nu$, as
\begin{equation}
\nu = \frac{v_0}{4 \sqrt{\rho}}N_t^2
\end{equation}
where $v_0$, $\rho$, and $N_t$ mean a typical velocity of patches, a number density of network cell, and a number density of patches 
respectively. We multiply $1/2$ taking the double counting into the account.
We obtained that the frequencies of merging and splitting are larger than that of cancellation in this study.
It can be deduced that the frequency distribution is maintained rapidly by merging and splitting compared to the timescale of cancellation.
This enables us to treat the number density of patches is time-independent in the evaluation of cancellation and apply the same analogy 
to the number density expanded in the dimension of flux content.
The collision frequency ,$\left. \partial N(\phi)/\partial t \right|_{\rm{col}}$, is evaluated as
\begin{equation}
\left. \frac{\partial N(\phi)}{\partial t} \right|_{\rm col}= - \frac{v_0 n_0^2 \phi_0^2}{4(\gamma -1)^2 \sqrt{\rho}} 
\left( \frac{\phi}{\phi_0} \right)^{-2\gamma+2}.
\end{equation}
Note that this total collision number becomes time-independent with an assumption of maintenance of a power-law flux distribution.
We assume that a total number of cancellation, 
$\left. \partial N(\phi)/\partial t \right|_{\rm{cnc}}$, equals to a total number of collision events of opposite polarities, 
$\left. \partial N(\phi)/\partial t \right|_{\rm{col}}$, namely
\begin{equation}
\left. \frac{\partial N(\phi)}{\partial t} \right|_{\rm{cnc}} = \left. \frac{\partial N(\phi)}{\partial t} \right|_{\rm{col}}.
\end{equation}
The frequency distribution of cancellation, $\left. \partial n(\phi)/\partial t \right|_{\rm{cnc}}$ is given by differentiating Eq.(16) with $\phi$, namely
\begin{equation}
\left. \frac{\partial n(\phi)}{\partial t} \right|_{\rm{cnc}}= 
\left. \frac{\partial ^2 N (\phi)}{\partial \phi \partial t} \right|_{\rm{cnc}} = 
- \frac{v_0 n_0^2 \phi_0}{2(\gamma-1) \sqrt{\rho}} \left( \frac{\phi}{\phi_0} \right)^{-2\gamma+1} .
\end{equation}
This assumption means that there is no patches passing through the patches of opposite polarity once they collide, which is difficult 
to check and we justify this assumption from the comparison of obtained frequency distribution of emergence in this model and that in the observation. 
We also evaluate a frequency distribution of emergence with assumptions of small amount of flux supply from the outside of the system
 and re-emergences of canceled fluxes.
These assumptions lead to the relationship that a frequency distribution of emergence nearly equates that of cancellation,
\begin{equation}
\left. \frac{\partial n(\phi)}{\partial t} \right|_{\rm{emrg}} \approx
- \left. \frac{\partial n(\phi)}{\partial t} \right|_{\rm{cnc}} =
\frac{v_0 n_0^2 \phi_0}{2(\gamma-1)\sqrt{\rho}} \left( \frac{\phi}{\phi_0} \right)^{-2\gamma+1}.
\end{equation}
We compare the power-law index and absolute value of frequency distribution of emergence in this model with the observational results.
Based on the above discussion our observation suggest a frequency distribution of emergence as 
$\left. \partial n / \partial t \right|_{\rm{emrg}} \thicksim 4.3 \times 10^{-35} \times \left( \phi / 1.0 
\times 10^{16} \, \rm{Mx} \right)^{-2.6}$ Mx$^{-1}$ cm$^{-2}$ s$^{-1}$
where we adopted $\rho = 1.0 \times 10^{-19} \ \rm{cm^{-2}}$ \citep{hag1997} and $v_0 = 2.0 \ \rm{km \  s^{-1}}$.
\cite{tho2011} reports a power-law frequency of emergence as 
$\left. \partial n / \partial t \right|_{\rm{emrg}} \thicksim 2.5 \times 10^{-35} \times \left( \phi / 
1.0 \times 10^{16} \, \rm{Mx} \right)^{-2.7}$ Mx$^{-1}$ cm$^{-2}$ s$^{-1}$.
The steepness of the power-law distribution and the absolute value in our model are in good agreement with the observational results.

We calculate flux replacement time by cancellation and emergence.
The total flux amount in the system is calculated from Eq.(13) as
\begin{equation}
\Phi_{\rm tot} = \int_{\phi_{\rm min}}^{\phi_{\rm max}} \phi^{\prime} \, n(\phi^{\prime}) \, d\phi^{\prime} 
\approx \frac{n_0 \phi_0^2}{2-\gamma} \left( \frac{\phi_{\rm max }}{\phi_0} \right)^{-\gamma+2}
\end{equation}
Since $\gamma < 2$, this result shows that the total flux is dominated by patches with larger flux content.
On the other hand,  we obtain a total flux loss amount by cancellation, $\left. \partial \Phi_{\rm tot }(\phi)/\partial t \right|_{\rm cnc}$, 
from Eq.(18) as 
\begin{equation}
\left. \frac{\partial \Phi_{\rm tot }}{\partial t} \right|_{\rm{cnc}} = 
\int_{\phi_{\rm{min}}}^{\phi_{\rm{max}}}\phi^{\prime} \left. \frac{\partial n(\phi^{\prime})}{\partial t} \right|_{\rm{cnc}} \, d\phi^{\prime}
\approx - \frac{v_0 n_0^2 \phi_0^3}{2 ( \gamma - 1)(2 \gamma - 3) \sqrt{\rho}} \left( \frac{\phi_{\rm{min}}}{\phi_0} \right)^{-2\gamma+3}.
\end{equation}
The total flux supply by recycled emergence is also evaluated in the same manner from Eq.(19) as 
\begin{equation}
\left. \frac{\partial \Phi_{\rm tot }}{\partial t} \right|_{\rm{emrg}}
=\int_{\phi_{\rm{min}}}^{\phi_{\rm{max}}}\phi^{\prime} \left. \frac{\partial n(\phi^{\prime})}{\partial t} \right|_{\rm{emrg}} \, d\phi^{\prime}
\approx \frac{v_0 n_0^2 \phi_0^3}{2 (\gamma - 1)(2 \gamma - 3) \sqrt{\rho}} \left( \frac{\phi_{\rm{min}}}{\phi_0} \right)^{-2\gamma+3}.
\end{equation}
Then flux replacement time is evaluated as
\begin{eqnarray}
\tau_{\rm{replace}}
& = &\Phi_{\rm tot}/ ( \left. \partial \Phi_{\rm tot}/\partial t \right|_{\rm cnc})
=\Phi_{\rm tot}/ ( \left. \partial \Phi_{\rm tot}/\partial t \right|_{\rm emrg}) \nonumber \\
& = &\frac{2 ( \gamma - 1 ) (2 \gamma - 3 )}{( 2 - \gamma ) v_0 n_0 \phi_0 \sqrt{\rho}} \left( \frac{\phi_{\rm{max}}}{\phi_0} \right)^{-\gamma+2}
\left( \frac{\phi_{\rm{min}}}{\phi_0} \right)^{2 \gamma-3}.
\end{eqnarray}
This result is qualitatively consistent with the previous result that flux replacement time becomes shorter with higher resolution \citep{mar1985, sch1998, hag2001}.

We summarize our interpretation from the discussion.
In our interpretation, a power-law frequency distribution of flux content is rapidly maintained by merging and splitting, which is supported 
by the result of a comparison of change rate of flux amount by each process.
Addition to this, we have found that a power-law frequency distribution with an index of $-2$ is a time-independent solution of splitting.
Cancellation is caused by convective dominant motion and frequency distribution of cancellation should naturally become a steep power-law distribution. 
Emergence is interpreted as re-emergence of submerged flux by cancellation, which is consistent with a steep frequency distribution of emergence 
\citep{tho2011}.
This should be one of the possibilities but the important point of this model is that the apparent flux transport through the photospheric layer becomes 
much more drastic when investigating emergence and cancellation although injected flux amount from the deeper layer is small or even zero. 
Recent paper, \cite{mey2011}, reports their numerical simulations of the magnetic carpet on the photosphere.
They show that a power-law distribution of flux content is maintained with an input of a steep power-law frequency of emergence.
The induced frequency of cancellation becomes a steep power-law one, which is consistent with our model.

What should be investigated for a further understanding of flux transport may be a statistical investigation of cancellation on the photosphere.
Another important question is whether there is a stable solution of our model with four magnetic processes or not.
We will investigate these questions in the series of this paper.

\acknowledgments

First of all, we want to represent our great appreciation to the referee, who read our manuscript 
carefully and gave a lot of useful advices.
We want to thank Global COE program From the Earth to $``$Earths$``$, which support author's stay at Lockheed Martin Solar and Astrophysics Laboratory. 
This work can not be done without this support.
We also extend our appreciation to the proofreading/editing assistance from the GCOE program. 
This work was supported by Grant-in-Aid for JSPS Fellows.
{\it Hinode} is a Japanese mission developed and launched by ISAS/JAXA, with NAOJ as domestic partner and NASA and STFC (UK) as
 international partners. It is operated by these agencies in co-operation with ESA and NSC (Norway).

\bibliographystyle{apj}
\bibliography{apj_iida_2011}

\begin{figure}
\epsscale{0.99}
\plotone{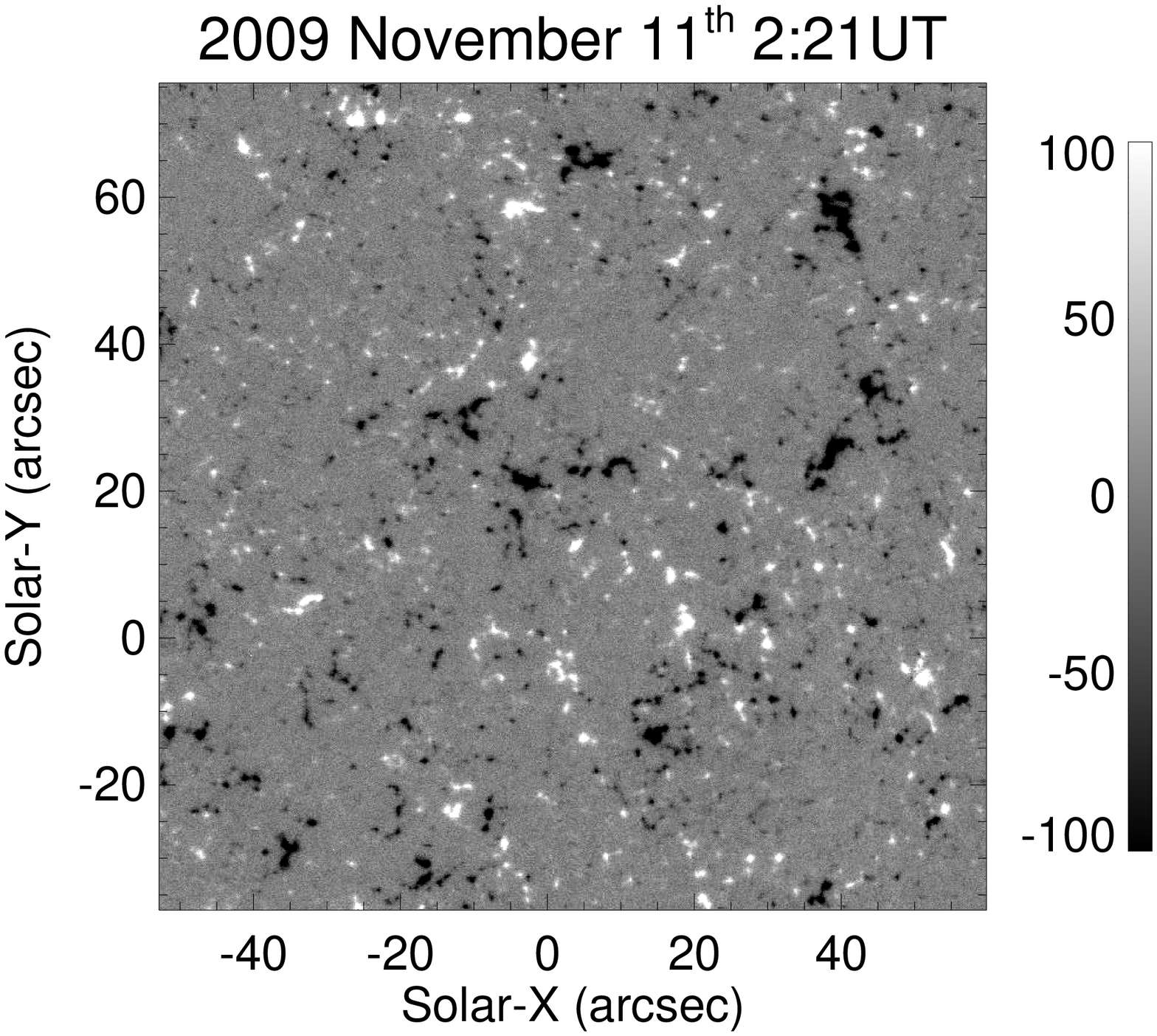}
\caption{An example of Na I D$_1$ magnetograms after the pre-process explained in the text. 
This magnetogram is taken at 2:21UT on 2009 November 11, which is the middle of observational period.
Some magnetic networks are contained in the field of view.}
\label{f1}
\end{figure}

\begin{figure}
\epsscale{0.8}
\plotone{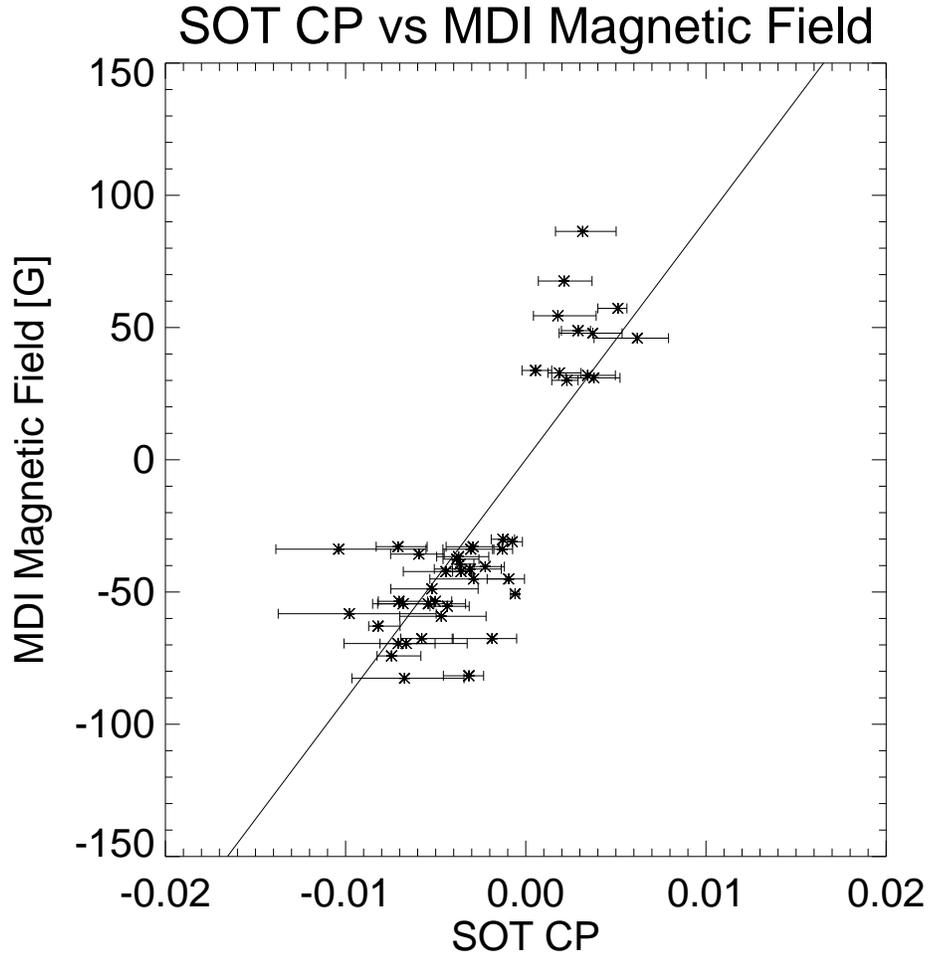}
\caption{A scatter plot of CP obtained by {\it Hinode}/SOT and magnetic flux obtained by SoHO/MDI.
Asterisks show averaged SOT CP signal corresponding to one pixel of MDI.
The bars in horizontal direction indicate minimum and maximum value of SOT CP signal in each MDI pixel.
Solid line shows a result of a linear fitting, whose slope is $9067.38$ G/DN.
The fitting range is from $30$ G to $100$ G in absolute magnetic field strength.}
\label{f2}
\end{figure}

\begin{figure}
\epsscale{0.99}
\plotone{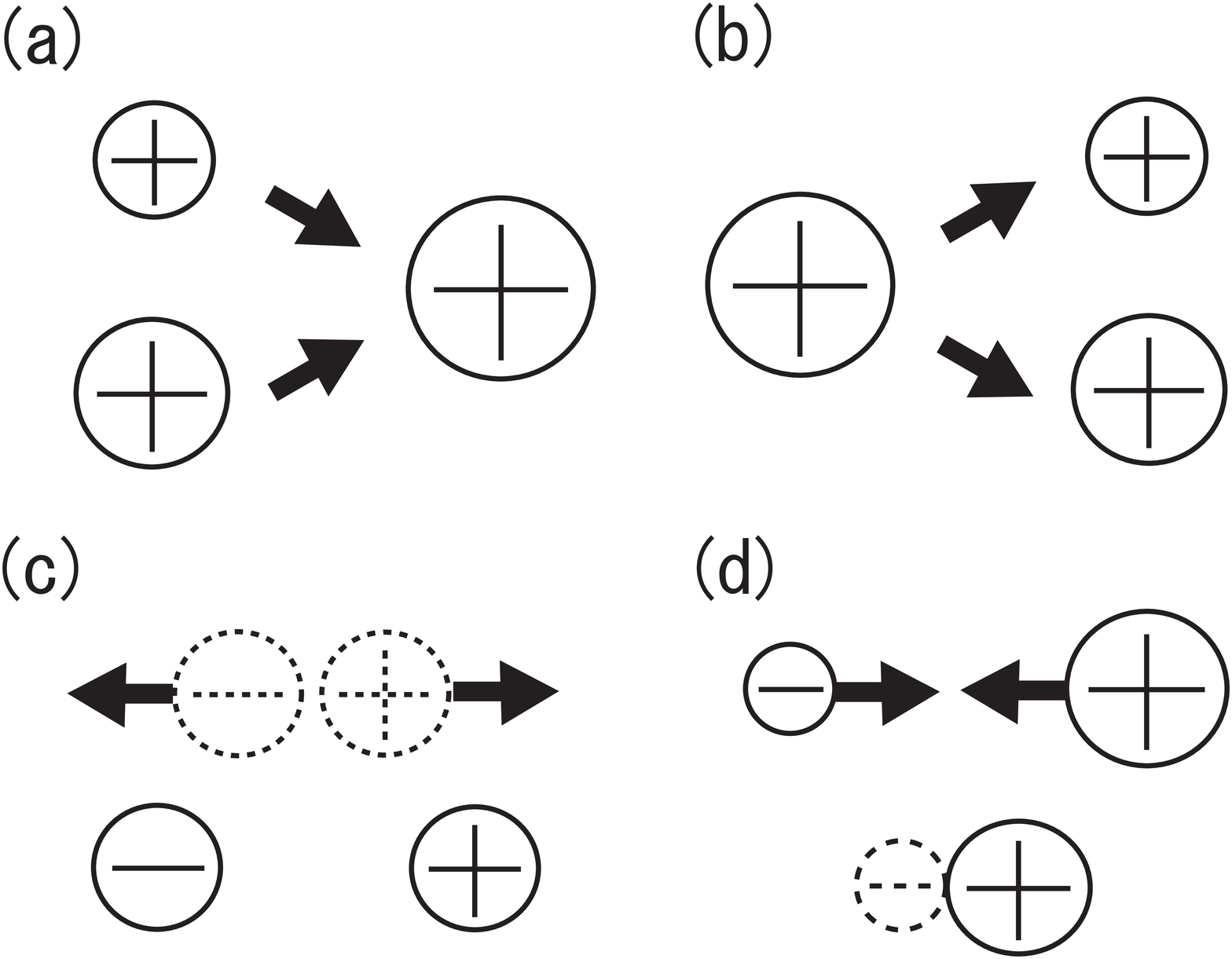}
\caption{Schematic pictures of four magnetic processes between two patches.
(a) Merging. Two patches converge and coalesce into one patch.
(b) Splitting. One patch is divided into two patches.
(c) Emergence. Two opposite polarities with equal amount of flux content appear.
(d) Cancellation. Two opposite polarities converge with each other. One of them disappears.
 The other loses its flux content or disappear.}
\label{f3}
\end{figure}

\begin{figure}
\epsscale{0.99}
\plotone{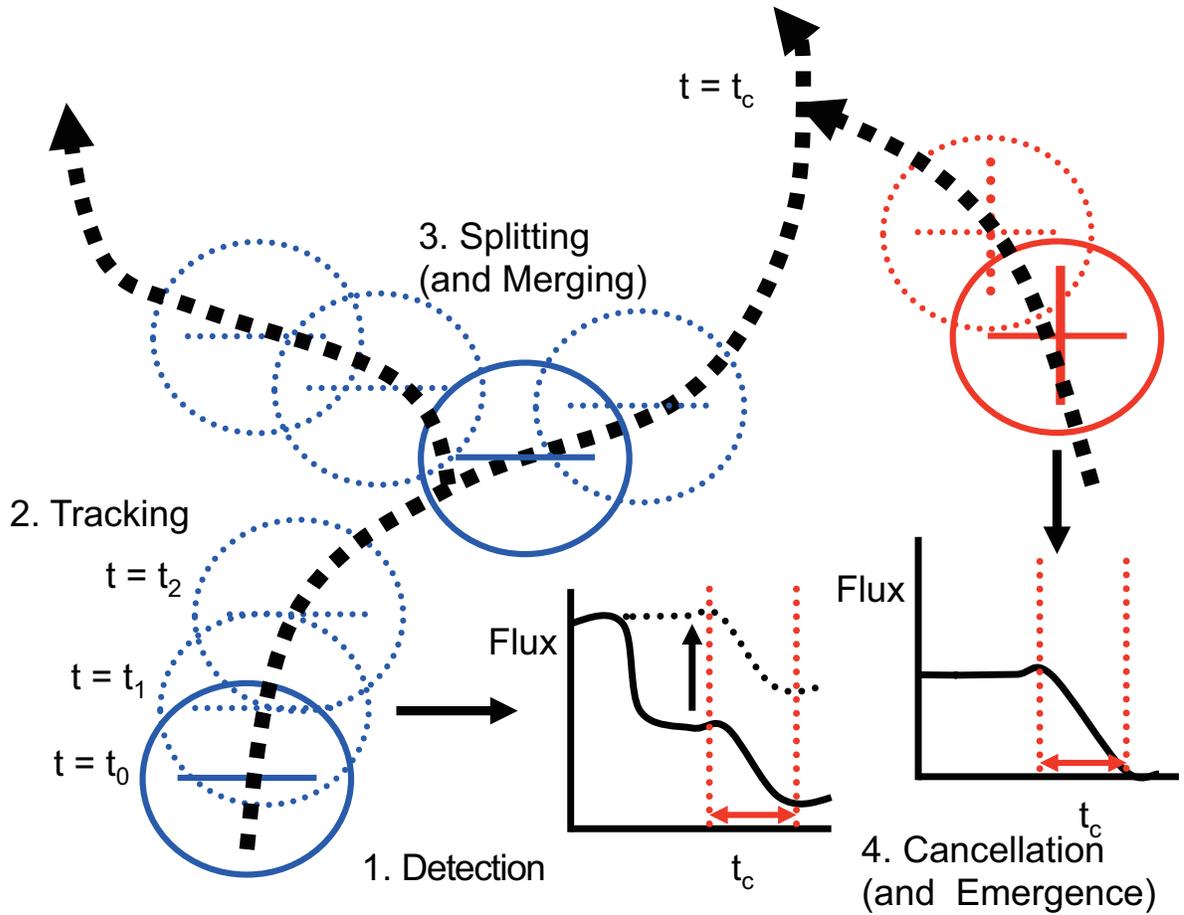}
\caption{
Schematic picture of our patch tracking and magnetic process detection. 
1. Detection of patches with clumping algorithm. 
2. Track of patches by examining overlaps. 
3. Detection of merging and splitting by examining overlaps. 
4. Detection of cancellation and emergence by making pairs of flux change events.
}
\label{f4}
\end{figure}

\begin{figure}
\epsscale{0.99}
\plotone{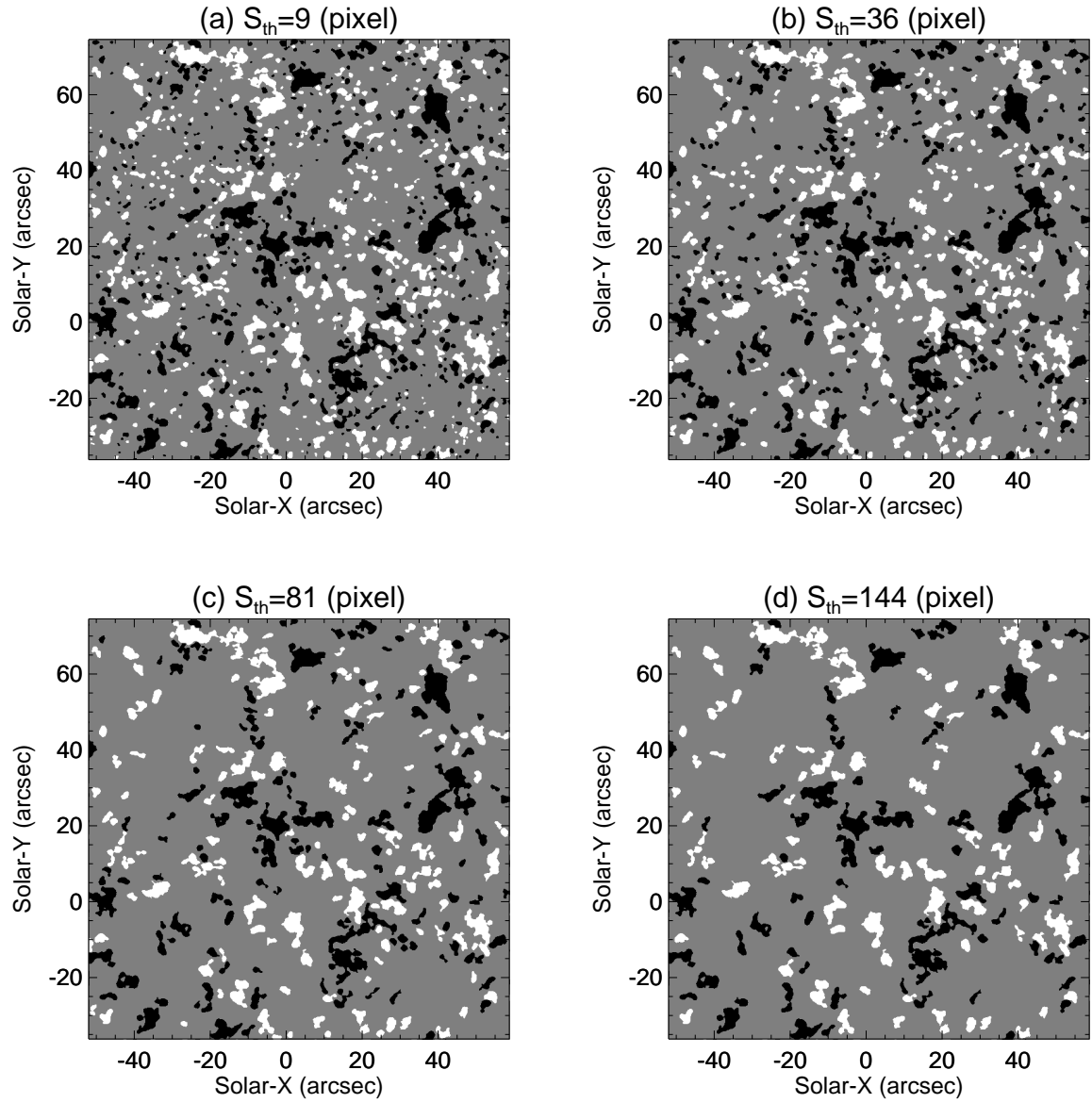}
\caption{
Two-leveled magnetograms taken at 2:21 on 2009 November 11 with different size thresholds. 
Size thresholds, $S_{\rm th}$, are set as (a) 9 pixels, (b) 36 pixels, (c) 81 pixels, and (d) 144 pixels.
}
\label{f5}
\end{figure}

\begin{figure}
\epsscale{0.99}
\plotone{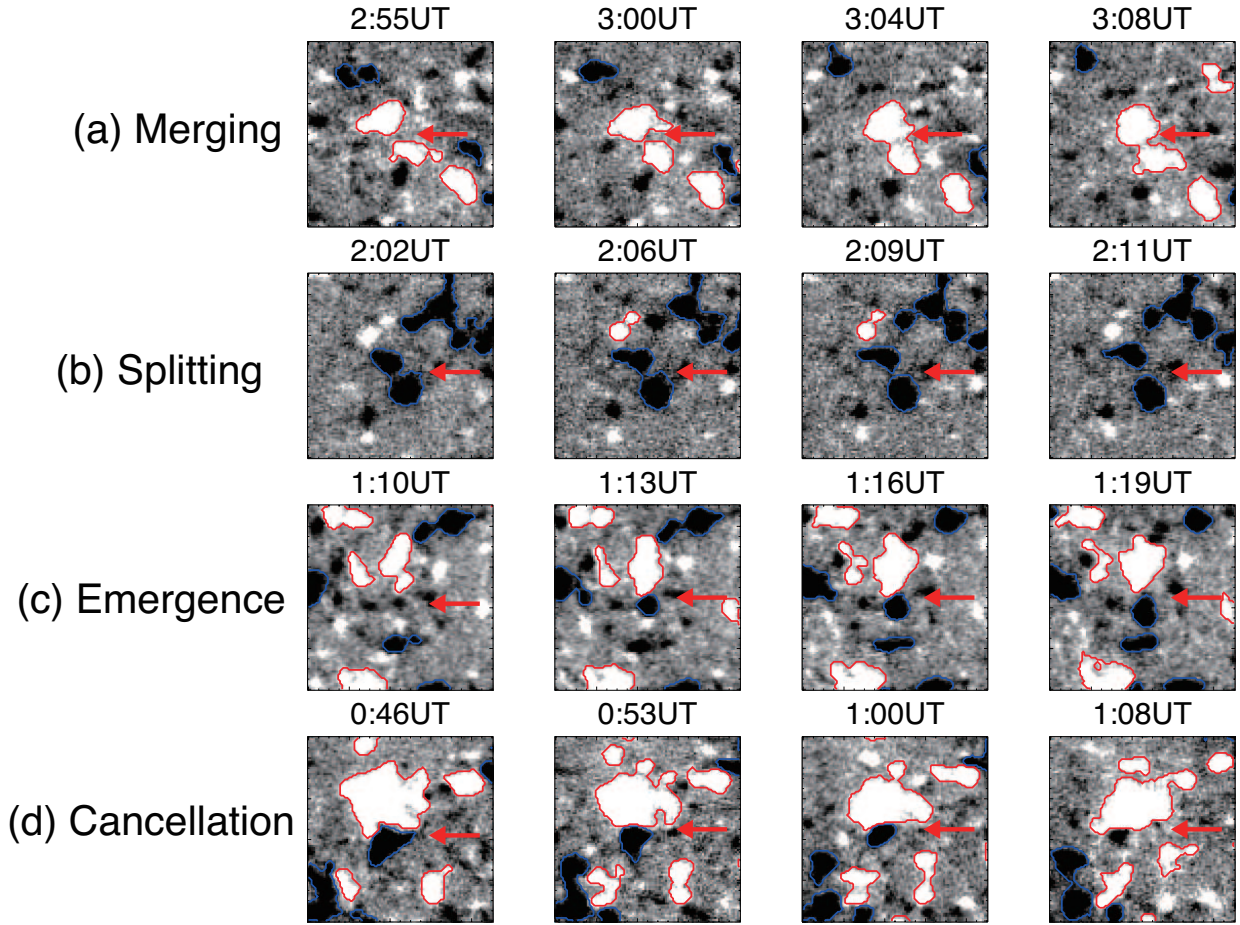}
\caption{
Examples of detected magnetic processes, namely (a) merging, (b) splitting, (c) emergence, and (d) cancellation. 
Background shows magnetic flux density obtained by {\it Hinode}/NFI. 
Red (blue) contours indicate positive (negative) patches detected with our threshold. The field of view is $14.4'' \times 14.4''$ for all the images.
}
\label{f6}
\end{figure}

\begin{figure}
\epsscale{0.99}
\plotone{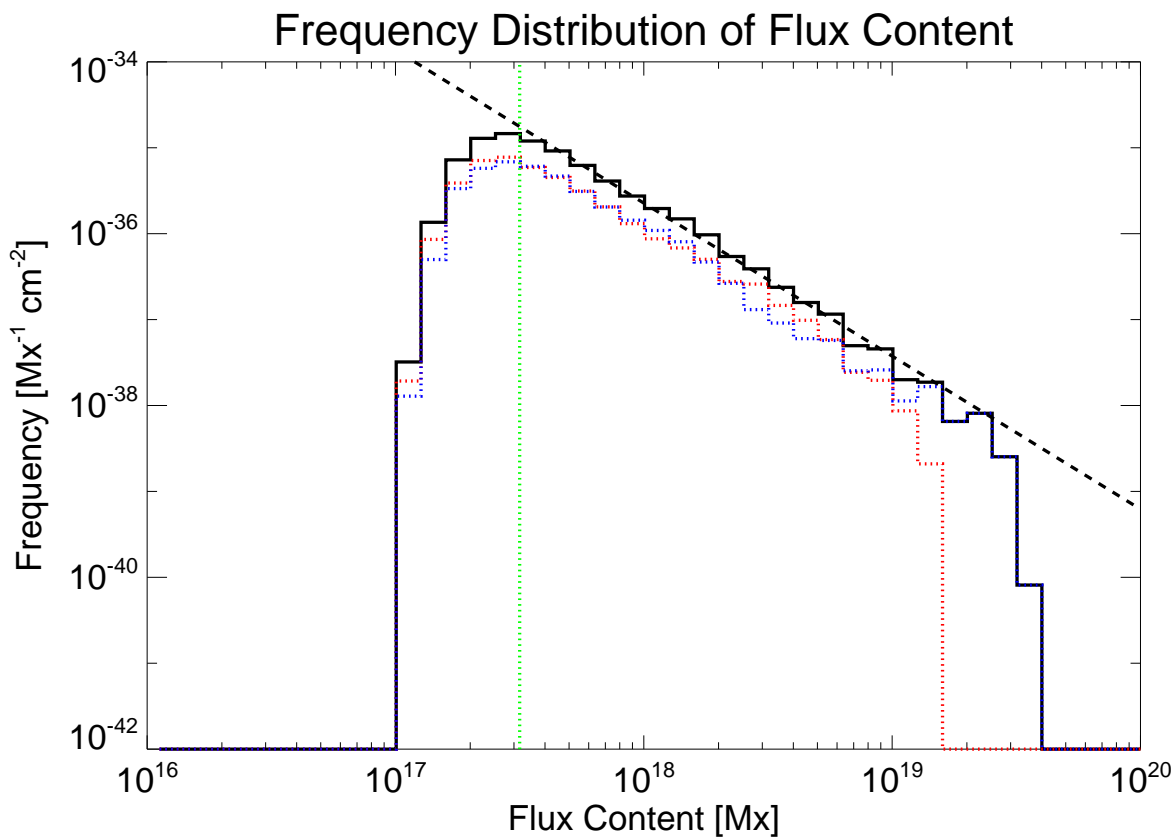}
\caption{
Frequency distribution of magnetic flux content. 
Red (blue) dashed histogram denotes that of positive (negative) polarity. 
Solid histogram denotes sum of them. Black dashed line indicates the fitting result of it. 
The obtained power-law index is -1.78 with a fitting range between 10$^{17.5}$ Mx and 10$^{19}$ Mx. 
The vertical green line denotes the detection limit, $\phi_{\rm th}$= 10$^{17.5}$ Mx.
}
\label{f7}
\end{figure}

\begin{figure}
\epsscale{0.99}
\plotone{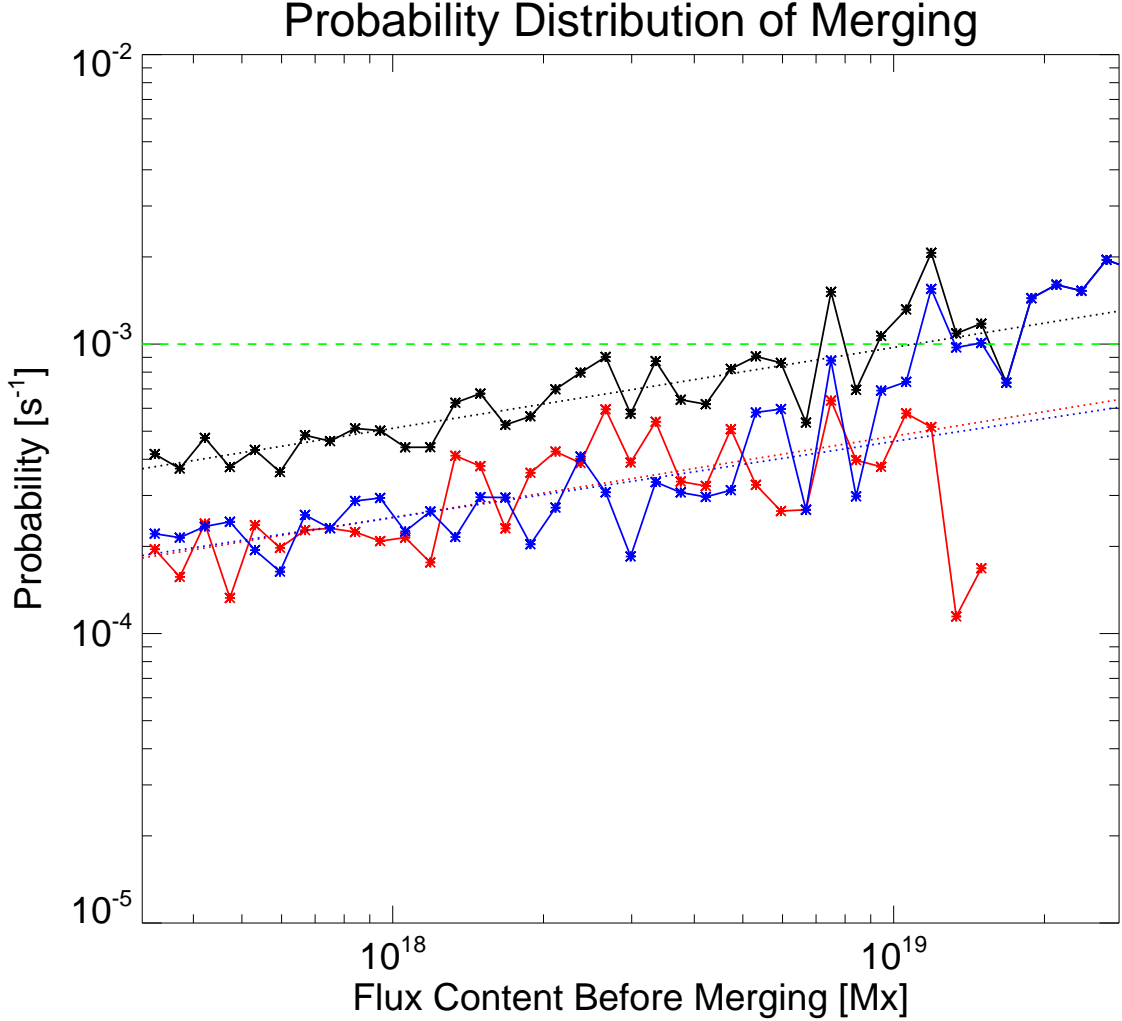}
\caption{
Apparent probability distribution of merging.
It is made from merging where the parent patches have flux content more than $\phi_{\rm th}$.
Red/blue/black solid lines indicate observational results for positive/negative/both patches.
Red/blue/black dashed lines indicate fitting results with a range of 10$^{17.5}$ - 10$^{19}$ Mx.
The power-law indexes of the fitting lines are 0.28, 0.26, and 0.28.
Horizontal and vertical dashed line indicates to timescale of 33 minutes and the detection limit, respectively.
}
\label{f8}
\end{figure}

\begin{figure}
\epsscale{0.99}
\plotone{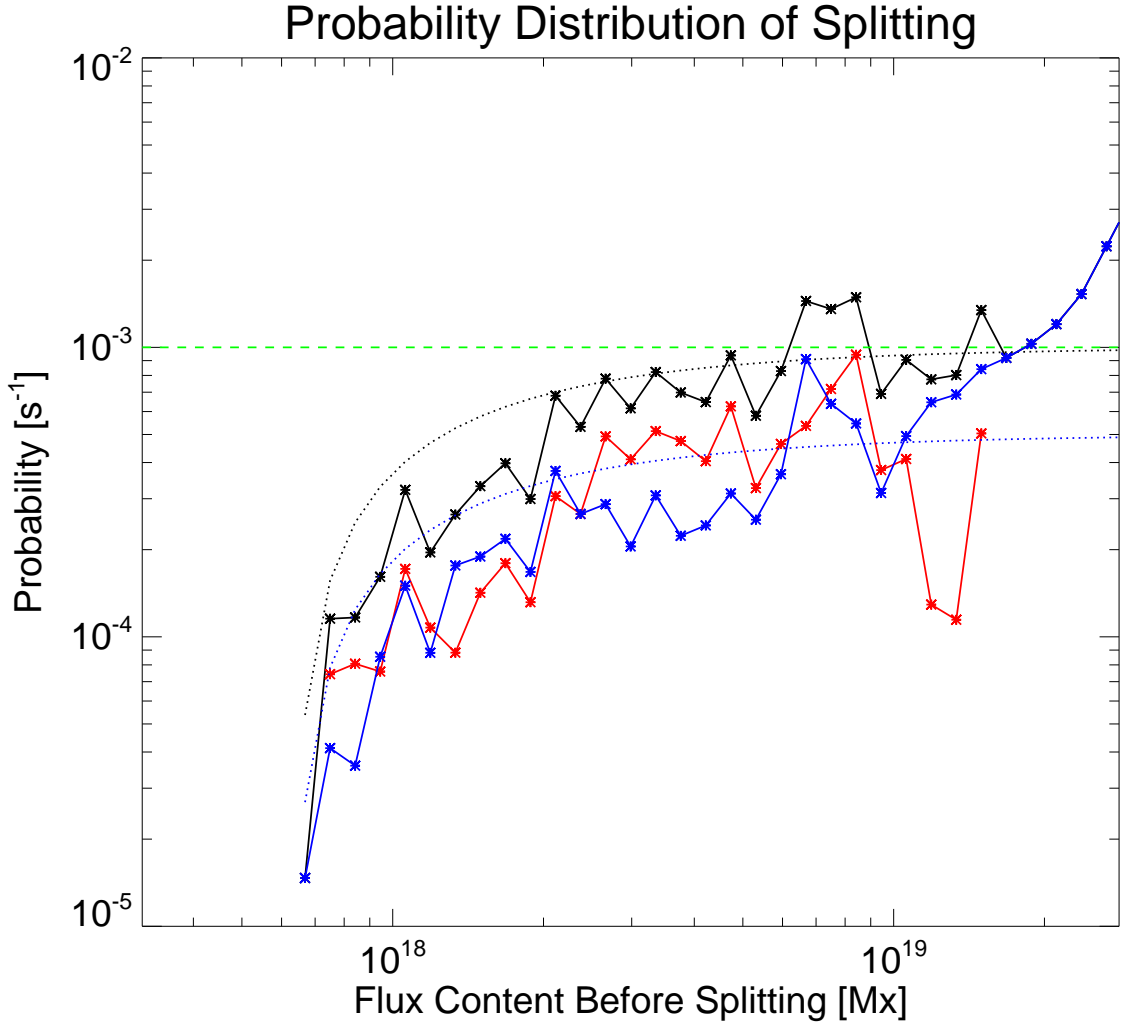}
\caption{
Apparent probability distribution of splitting.
Red/blue/black solid lines indicate observational results for posi- tive/negative/both patches.
Blue and black dashed lines indicate analytical curves with $k_0$ = 5.0$\times$10$^{-4}$ s$^{-1}$ and 1.0$\times$10$^{-3}$ s$^{-1}$. 
Horizontal dashed line indicates a timescale of 33 minutes.
}
\label{f9}
\end{figure}

\begin{figure}
\epsscale{0.99}
\plotone{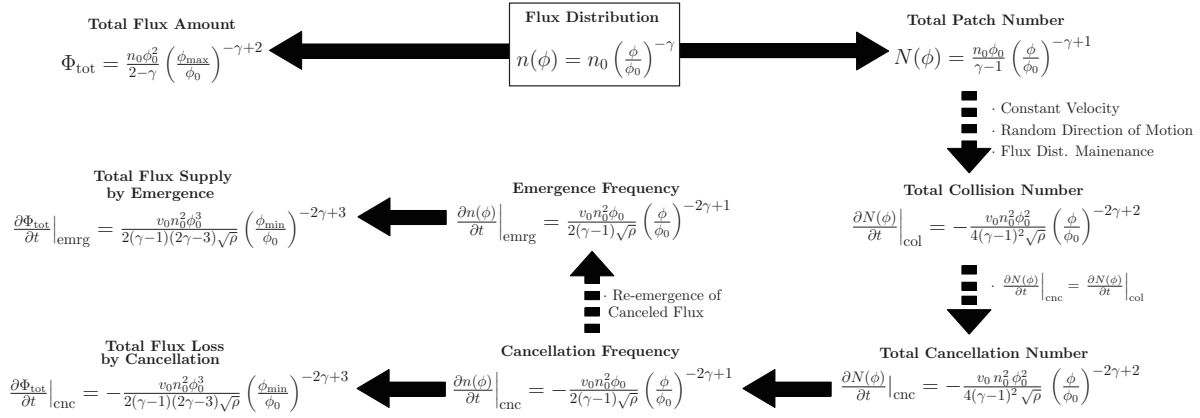}
\caption{Model of relationship among flux distributions of flux content, cancellation, and emergence with $1.5<\gamma<2$.
Solid arrows indicate mathematical relations, while dashed arrows indicate relationships with some assumptions.
}
\label{f10}
\end{figure}

\begin{table}
\begin{center}
\begin{tabular}{l|c|c}
\hline\hline
&Positive Polarity&Negative Polarity\\
\hline
Patch&1636 (2.53 Mx cm$^{-2}$)&1637 (3.60 Mx cm$^{-2}$)\\
Merging&536 (1.65 $\times$ 10$^{-3}$ Mx cm$^{-2}$ s$^{-1}$)&535 (3.03 $\times$ 10$^{-3}$ Mx cm$^{-2}$ s$^{-1}$)\\
Splitting&493 (1.48 $\times$ 10$^{-3}$ Mx cm$^{-2}$ s$^{-1}$)&482 (3.53 $\times$ 10$^{-3}$ Mx cm$^{-2}$ s$^{-1}$)\\
\hline
Emergence&\multicolumn{2}{c}{3}\\
Cancellation&\multicolumn{2}{c}{86}\\
\hline\hline
\end{tabular}
\end{center}
\caption{Total numbers and flux amounts of magnetic patches and magnetic processes.
The numbers in the parentheses in the first line are averaged flux densities of each polarity.
Those in the parentheses in the second and third lines are averaged rates of flux density change by each process.
These values are averaged over the whole field of view and observational period.
}
\label{tab1}
\end{table}

\end{document}